# Delayed Choice Quantum Eraser Experiment Revisited: Causality and Informational Coherence

Taku Ohwada

Department of Aeronautics and Astronautics, Graduate School of Engineering
Kyoto University, Kyoto 615-8540, JAPAN

**Abstract**
An operationally well-defined delayed-choice quantum-eraser experiment is proposed, realizing a genuine delayed choice within presently available quantum-optical technology. A multimode quantum memory supplies a controlled and verifiable delay, ensuring that the choice operation is applied strictly after the observation event. Electronic single-photon interference detection measurements furnish a direct statistical discriminator between the causal and informational coherence hypotheses, based solely on marginal detection statistics and without any post-selection.

## 1. Introduction

Quantum mechanics has a long history of challenging human intuition. Phenomena such as wave–particle duality, wavefunction collapse, and quantum entanglement have each defied our classical understanding of nature, raising fundamental questions about physical reality. These concepts, though counterintuitive, have been fully integrated into quantum mechanics, supported by overwhelming experimental evidence and strong formal consistency. Nevertheless, one foundational principle in physics at large has been kept intact throughout the development of quantum mechanics: causality. Regardless of how nonlocal quantum correlations may appear, the core temporal structure—the past influencing the future, but never the reverse—has been strictly upheld to date.

Wheeler's delayed-choice experiment [1] was originally proposed as a thought experiment intended to illustrate how easily our classical intuition can fail when applied to quantum phenomena. In interferometric settings, it may seem as though a later choice—such as whether to allow or suppress interference—retroactively affects an earlier event. In Wheeler's original scenario, however, the choice that distinguishes interference from non-interference is always made before detection. Consequently, no genuine reversal of temporal order occurs, and the result remains fully consistent with standard quantum mechanics and therefore with causality.

The delayed-choice quantum eraser (DCQE) [2–3] appears, at first sight, to go one step further. In these experiments, the choice associated with erasing or preserving which-path information is implemented after detection. The interference patterns observed in DCQE arise only through post-selection, typically via coincidence counting with idler photons. The apparent retroactive effect is considered to originate not from the delayed choice itself but from a filtering procedure applied afterward to the recorded data. For this reason,

DCQE experiments are widely recognized as consistent with quantum mechanics, despite their strikingly counterintuitive appearance.

In most experimental realizations of DCQE based on spontaneous parametric down-conversion (SPDC), the absence of single-photon self-interference on the signal side is commonly attributed to the presence of entanglement. However, this interpretation conflates entanglement itself with the mixed nature of the reduced signal state that typically arises from strong spectral and spatial correlations inherent in standard SPDC sources. In such configurations, the signal photon fails to exhibit self-interference not because it is entangled, but because its reduced density matrix lacks sufficient purity. In the present work, we explicitly separate these two notions and require that the signal photon be prepared in a state capable of single-photon self-interference independently of any delayed operation applied to the idler.

There are two reasons why existing DCQE experiments are not appropriate as genuine tests of causality. First, the time at which an experimental outcome is finalized is not fixed at the moment of detection. Instead, the outcome is declared only after subsequent data processing, so that the effective "observation" can, in principle, be postponed arbitrarily late—after lunch or even after returning from a Caribbean vacation—without altering the reported result. Second, the reported interference relies on discarding a substantial portion of the recorded data. From an engineering and signal-processing viewpoint, structures obtained through such data rejection cannot appropriately be regarded as physical phenomena. If it were allowed, one could equally extract structured signals—for example, a whistle—from pure white noise simply by filtering and discarding data.

Meanwhile, rapid advances in quantum information science have elevated informational coherence to a central role in the quantum framework. In particular, the consistency of information within entangled systems has become essential to quantum computing, quantum communication, and quantum error correction. In these settings, informational coherence is no longer an interpretational notion, but is operationally defined and tested through concrete experimental protocols.

In a genuine DCQE configuration without post-selection or coincidence filtering, where the signal photon is detected and its outcome becomes informationally fixed before any manipulation of the idler photon, the simultaneous maintenance of causal ordering and informational coherence becomes nontrivial. In such a setting, the delayed erasure operation can no longer be absorbed into conditional sampling or post-selection. Depending on the experimental outcome, either the marginal detection statistics of the signal photons remain independent of the delayed choice, or they exhibit a statistically detectable dependence on that choice. The two principles therefore lead to operationally distinct predictions once the strict temporal separation between the detection event and the choice event is satisfied, a condition referred to in this paper as Wheeler's condition.

This paper proposes an experimental architecture that provides an operational test of the compatibility between two principles: causality and informational coherence. By introducing a multimode quantum memory on the idler side, we enforce strict temporal separation between the completion of the detection of the signal photons and the application of the choice operation on the idler. The quantum memory serves solely as a timing element and does not introduce any additional physical assumptions or dynamical effects. This setup enforces Wheeler's condition, in which the "choice" is made only after the detection outcome is finalized. The proposed scheme enables a genuine DCQE configuration without post-selection or coincidence filtering.

## 2. Operational Criterion for Causality

### 2.1 Operational Definition of Observation

In order to formulate a genuine DCQE experiment, it is essential to specify precisely when an experimental outcome is regarded as finalized. In the present work, we adopt a strictly operational viewpoint and distinguish between the completion of a physical detection event and the completion of an observation. Here, we regard an observation as completed only when the detection outcomes are finalized as operationally well-defined statistical data, according to a predetermined protocol.

We define $T_{\text{phys}}$ as the time at which the relevant physical detection process is completed. This corresponds to an irreversible physical event at the detector, such as the generation of an avalanche signal in a single-photon detector, accompanied by macroscopic energy dissipation. After $T_{\text{phys}}$, no further physical processes capable of affecting the detection outcomes remain. We define $T_{\text{observe}}$ as the time at which the detection outcome is operationally fixed as an observation. The observation time $T_{\text{observe}}$ may exceed $T_{\text{phys}}$ by a finite margin, accounting for electronic response, signal registration, and stabilization of the detection record. In practice, one may write

$$T_{\text{observe}} = (1 + \epsilon)\, T_{\text{phys}},$$

with a finite safety margin $\epsilon > 0$, emphasizing that observation is finalized only after the physical detection process has irreversibly completed.

This distinction is crucial for the present discussion. In particular, $T_{\text{observe}}$ is defined independently of any subsequent data processing, post-selection, or coincidence analysis. Once $T_{\text{observe}}$ is reached, the observation outcome is considered finalized and cannot be altered or reinterpreted by later operations performed elsewhere in the experimental apparatus. Experiments in which $T_{\text{observe}}$ cannot be operationally defined in this manner— namely, experiments where the outcome is declared only after conditional sampling or retrospective data filtering—are excluded from consideration in the present framework.

### 2.2 Wheeler's Condition

Let $T_{\text{choice}}$ denote the time at which the collective delayed–erasure operation starts, i.e., the moment when the decision to change the polarization states of the ensemble of idler photons is physically implemented. For a configuration to qualify as a genuine delayed–choice scenario, this operation must take place *strictly after* $T_{\text{observe}}$, ensuring that the measurement outcomes of the signal photons are already fixed before the choice is made. Thus, a necessary condition for any genuine test of causality in DCQE configurations can be formalized as:

$$T_{\text{observe}} < T_{\text{choice}}.$$

We refer to this inequality as Wheeler's condition. It specifies the temporal ordering required for a genuine delayed-choice configuration. In the following, we proceed to propose an experimental configuration that satisfies Wheeler's condition by design.

## 3. Sketch of Experiment

As in conventional DCQE experiments, our proposal employs entangled photon pairs generated via SPDC. Each generated pair consists of two photons, conventionally referred to as the signal and the idler. The signal photon is routed into a detection subsystem, while the idler photon is sent toward a delayed erasure operation. The signal path is split into two arms by a 50:50 fiber beam splitter, forming an interferometric setup. Path distinguishability is introduced by encoding which–path information into polarization: the two paths are assigned mutually orthogonal polarization states, thereby suppressing interference unless the which–path information is erased. A subsequent polarization–erasure operation applied to the delayed idler photons determines whether the which–path information assigned to the signal photons is preserved or removed.

The crucial feature of our proposal is the inclusion of a multimode quantum–memory device, placed between the SPDC source and the polarization–control unit on the idler side. This guarantees strict temporal separation between signal detection and idler erasure, thereby enforcing Wheeler's condition by design.

### 3.1 Delayed Choice: Experimental Wheeler's Condition
The introduction of a multimode quantum memory allows the delayed–choice operation to be separated from signal-photon detection by a controllable temporal interval determined by the memory storage time. To formalize this separation, we distinguish three characteristic timescales.

Let $T_{\text{choice}}$ denote the time at which the collective idler-state operation—preserving or erasing the which–path information—is executed. Wheeler's condition requires

$$T_{\text{observe}} < T_{\text{choice}},$$

so that the delayed choice is made only after the signal detection outcome has become informationally fixed.

The quantum memory defines a delay time $T_{\text{delay}}$ such that, for any time $t < T_{\text{delay}}$, none of the idler photons has yet passed through the which-path controller. For times $t \geq T_{\text{delay}}$, this condition is no longer satisfied, and the delayed-choice operation is therefore not operationally defined. The delayed operation must therefore be executed while all the idlers are still within the memory, which imposes the requirement

$$T_{\text{choice}} < T_{\text{delay}}.$$

Combining this inequality with Wheeler's condition, the operational ordering required for a genuine delayed-choice configuration is expressed as

$$T_{\text{observe}} < T_{\text{choice}} < T_{\text{delay}}.$$

This inequality establishes the operational ordering needed for a genuine delayed–choice erasure: the signal outcome must be finalized before the choice is made, and the idler states must remain coherently stored until that choice is implemented.

### 3.2 Operational Unit, Memory, and SPDC

Current multimode quantum memories can typically offer either long storage times ($\gtrsim 10^2$ $\mu s$) or large mode capacities ($\sim 10^3$), while combining both capabilities in a single device remains an ongoing development target. Nevertheless, a memory providing several hundred microseconds of storage together with several thousand temporal modes is widely regarded as a near-term milestone on the path toward scalable quantum information processing [4,5].

In the present proposal, we assume an idler–storage delay of

$$T_{\text{delay}} = 500 \ \mu s,$$

and a signal–photon acquisition window of

$$T_{\text{phys}} = 400 \ \mu s,$$

which together ensure strict temporal separation between the signal detection and any delayed operation applied to the idler photons, with a finite safety margin.

Each operational unit consists of approximately $N = 500$ signal–idler pairs. To suppress multiple occupancy of temporal bins, the memory must provide a number of modes substantially larger than the number of stored photons. According to the standard conservative choice of three temporal modes per photon, commonly adopted to suppress mode overlap, we require about

$$3N = 1500$$

temporal modes as the memory capacity. Given the $400 \ \mu s$ acquisition window for the 500 idler photons, this requirement immediately yields a target temporal–bin resolution

$$\Delta t_{\text{bin}} = \frac{400 \, \mu s}{3 \times 500} \sim 266 \, ns,$$

which is well within the capabilities of current multimode quantum memories, the minimum resolution of which typically lies in the $10 - 20 \, ns$ range.

Finally, the SPDC source must supply the required $\sim 500$ signal–idler pairs within the $400 \, \mu s$ acquisition window. Using standard parameter values

$$P_S = P_I = P = 0.3, \quad p = 0.01,$$

where $P_S$, $P_I$, and $p$ denote the signal–collection efficiency, idler–collection efficiency, and per–pulse pair–generation probability, the required pump repetition rate is

$$f_{\text{pump}} = \frac{500}{400 \, \mu s} \times \frac{1}{P} \times \frac{1}{p} = 4.17 \times 10^8 \, \text{Hz} = 417 \, \text{MHz}.$$

Such sub–gigahertz pump rates are entirely routine in contemporary SPDC platforms.

The choice $p = 0.01$ also suppresses double occupancy of a temporal bin, since multi–pair generation in a single pump pulse occurs only with probability $O(p^2) \sim 10^{-4}$. Among the $\mu_0 = N/P \sim 1700$ entangled pairs originally produced by SPDC in each operational unit, $\mu_0 P^2 = NP \sim 150$ pairs are jointly collected. Thus, approximately 150 of the 500 signal photons are entangled with idler photons stored in the memory.

We further assume that the SPDC source is engineered such that the reduced state of each signal photon is sufficiently pure to support single-photon self-interference, for example by suppressing strong spectral and spatial correlations while retaining entanglement in an auxiliary degree of freedom.

### 3.3 Electronic Interference Detection

Suppose that which–path information is erased for the entangled subset of 150 signal photons whose idler partners are stored in the memory, while it remains preserved for the remaining 350 photons. Under these conditions, only the entangled subset carries the potential to exhibit interference. However, a sample size of 150 photons is clearly insufficient to produce any visible interference pattern on a screen. Spatial detection schemes relying on aggregate fringe visibility are therefore impractical in the present context.

We thus adopt an alternative approach in which the presence or absence of interference is inferred directly from single–photon detection statistics in an electronic measurement scheme. In this configuration, interference manifests as a genuinely single–photon phenomenon: each signal photon interferes solely with the two components of its own amplitude after passing through a 50:50 fiber splitter. For this interpretation to remain valid, two conditions must be satisfied: (i) different signal photons must not arrive within each other's coherence windows, and (ii) multi–pair SPDC emissions must be negligible, ensuring that no second photon contributes amplitude to the interference process.

Both requirements are quantitatively satisfied under the parameters specified in Sec. 3.2. The temporal spacing between successive signal photons greatly exceeds their coherence times, and the chosen pump repetition rate and pair–generation probability imply that multi–pair events occur only with probability $O(p^2) \sim 10^{-4}$. Consequently, the observed interference can be attributed unambiguously to the self–interference of individual signal photons.

The interferometric arrangement—splitting a single photon into two spatial modes and encoding which–path information via orthogonal polarization—is standard in quantum optics. The interference is inferred from the detection rate at the monitored output port of the final coupler:

- *With which–path information erased*, all interfering photons exit through the monitored port with probability 1.
- *With which–path information preserved*, the two output ports register photons with equal probability, yielding a detection probability of 1/2 at the monitored port.

The presence of interference is thus operationally signaled by a statistical shift in detection probability from 0.5 to 1.

For each operational unit, three terminal ports are always terminated by standard absorbers:

(i) the unused output of the final coupler,

(ii) the signal arm after detection, and

(iii) the idler arm after the polarization operation.

Routine suppression of stray light around the memory I/O is also assumed. Meanwhile, the idler readout from the memory is completed within a few microseconds ($\lesssim 10$ μs), and all signal/idler ports are promptly terminated by standard absorbers, so that each operational unit is fully closed temporally at $T_{\text{end}} \sim 510$ μs. Thus, the optical geometry of each unit is fully local and temporally closed. We emphasize that the electronic detection records only the marginal single-photon statistics and not the post-collapse spatial distribution.

## 3.4 Statistical Perspective

We analyze the stochastic behavior of the detection events within the framework of Poisson statistics, which provides a structurally transparent description of the successive filtering processes inherent in the setup. Each physical stage—pair generation, signal collection, idler survival, and the final interferometric routing—acts as an independent thinning applied to an underlying Poisson stream.

(1) Pair generation: the mother Poisson process
The SPDC source produces photon pairs independently across pump pulses. Over the acquisition window $T_{\text{window}}$, the number of generated pairs $N_{\text{pair}}$ follows

$$N_{\text{pair}} \sim \text{Poisson}(\mu_0), \qquad \mu_0 = f_{\text{pump}}\, p\, T_{\text{window}} \sim 1700,$$

which serves as the mother process from which all subsequent thinning operations originate.

(2) Signal-side thinning
A generated pair contributes a usable signal photon with probability $P_S$. Thus, the flux of signal photons entering the interferometric stage is obtained by the thinning

$$\mu_0 \longrightarrow \mu_0 P_S.$$

Up to this point, both the causality-preserving and informational-coherence models share the same probabilistic structure.

(3) Causality-preserving model: thinning stops at the coupler
Under the causality-preserving hypothesis, the delayed operation on the idler photon has no influence on the detection statistics of the already recorded signal photon. No further informational filtering acts on the signal stream. Each signal photon emerging from the final 50:50 coupler reaches the monitored port with probability $1/2$, producing the thinning chain

$$\mu_0 \longrightarrow \mu_0 P_S \longrightarrow \frac{\mu_0 P_S}{2}.$$

Thus, the detection statistics follow

$$N_{\text{det}}^{(\text{causal})} \sim \text{Poisson}\left(\frac{1}{2}\mu_0 P_S\right),$$

with mean and standard deviation

$$\mathbb{E}[N_{\text{det}}^{(\text{causal})}] = \frac{\mu_0 P_S}{2}, \qquad \sigma_{\text{causal}} = \sqrt{\frac{\mu_0 P_S}{2}}.$$

(4) Informational–coherence model
If informational coherence holds, the delayed erasure operation determines a subset of signal photons whose idler partners are both collected and retain sufficient entanglement fidelity. A generated pair is therefore interference-capable with probability $P_I F$, while the complementary fraction $1 - P_I F$ remains which-path–preserving and cannot contribute to interference. This yields the deeper thinning hierarchy

$$\mu_0 \to \mu_0 P_S \to \begin{cases} \mu_0 P_S P_I F & \text{(interference-capable)}, \\ \mu_0 P_S (1 - P_I F) & \text{(which-path–preserving)}. \end{cases}$$

At the final coupler, the interference-capable component reaches the monitored port with probability 1, whereas the which-path–preserving component contributes with probability 1/2. Hence the total detection flux becomes

$$\mu_{\text{det}}^{(IC)} = \mu_0 P_S P_I F + \frac{\mu_0 P_S (1 - P_I F)}{2} = \frac{\mu_0 P_S (1 + P_I F)}{2},$$

and the detection statistics follow

$$N_{\text{det}}^{(IC)} \sim \text{Poisson}\left(\frac{\mu_0 P_S (1 + P_I F)}{2}\right).$$

with mean and standard deviation:

$$\mathbb{E}\left[N_{\text{det}}^{(IC)}\right] = \frac{\mu_0 P_S (1 + P_I F)}{2}, \qquad \sigma_{\text{IC}} = \sqrt{\frac{\mu_0 P_S (1 + P_I F)}{2}}.$$

## 4. Discussions and Implications

### 4.1 Summary of the Proposal

We have proposed an operationally well-defined delayed-choice quantum eraser experiment that enables a direct experimental test of the compatibility between two foundational principles: causality and informational coherence. A central contribution of the present work is that both principles are cast into an explicitly operational form, allowing their consequences to be tested at the level of experimentally finalized detection statistics.

By fixing the temporal ordering between observation and choice through a multimode quantum memory, the proposed scheme establishes a concrete notion of observation time $T_{\text{observe}}$. Within this framework, the delayed erasure operation cannot be absorbed into post-selection or conditional data analysis, and the status of causal ordering becomes experimentally testable rather than interpretational.

As a result, causality-preserving and informational-coherence descriptions lead to distinct and measurable predictions for the marginal detection statistics of the signal photons once Wheeler's condition $T_{\text{observe}} < T_{\text{choice}}$ is satisfied. The experiment therefore provides a rare example in which principles that are often discussed at a conceptual or interpretational level are brought into direct operational confrontation.

### 4.2 Variant Experiments

The experimental scheme proposed here admits simple implementation variants arising from practical constraints on quantum-memory capacity. When the number of available memory modes is limited, the experiment may be realized either by distributing identical experimental units in parallel or by reusing the same unit sequentially in time. In both variants, the final observable is obtained by aggregating marginal detection statistics and is identical to that of the original scheme, provided Wheeler's condition is satisfied. This equivalence follows because, in both the causality-preserving and informational-coherence models, the marginal detection statistics are independent of the temporal distribution of signal-photon detection events.

### 4.3 Possible Experimental Outcomes

Depending on the observed statistics, the experiment admits three qualitatively distinct regimes:

(i) Statistics consistent with causal independence.
The delayed choice produces no statistically detectable influence on the marginal detection statistics of the signal photons. This outcome is consistent with the standard quantum-mechanical expectation that later operations do not affect earlier marginal outcomes. Importantly, within the present framework, this regime also constrains the operational reach of informational coherence: although entanglement enforces strong correlations, its informational consistency does not extend backward in time to alter already finalized observation statistics when Wheeler's condition is strictly satisfied. In this sense, the result constitutes a nontrivial null test, placing an explicit operational bound on how informational coherence and causal ordering coexist.

(ii) Statistics not captured by either reference model.
The observed statistics may fall outside the confidence intervals predicted by both the causality-preserving model and the informational-coherence model. Such a result would indicate that neither framework provides an adequate description of the observed behavior. Rather than being a failure mode, this regime would signal the presence of previously unaccounted-for structure—either experimental or theoretical—in the interplay between temporal ordering and information flow. Identifying the origin of such deviations would constitute a concrete target for further theoretical refinement and experimental investigation.

(iii) Statistics exhibiting dependence on the delayed choice.
The marginal detection statistics of the signal photons exhibit a statistically significant dependence on the delayed erasure operation, despite strict satisfaction of Wheeler's condition. This regime represents a direct operational tension between causal ordering and informational coherence: the statistics of an already finalized observation correlate with a later choice. Such an outcome would go beyond interpretational ambiguity and would demand a reassessment of how information, entanglement, and temporal ordering are jointly treated within quantum theory.